\documentclass[12pt,a4paper]{article}

\usepackage{graphics}
\usepackage{latexsym}
\usepackage{amsmath}
\usepackage{amssymb}

\title{Global warming: What does the data tell us?}
\author{E. X. Alb\'{a}n and B. Hoeneisen}
\date{\small{Universidad San Francisco de Quito \\
	19 February 2002}}
\begin{document}
\maketitle

\begin{abstract}
\noindent
We analyze global surface temperature data obtained
at $13472$ weather stations from the year 1702 to 1990.
The mean annual temperature of a station
fluctuates from year to year by typically
$\approx \pm0.6^o$C (one standard deviation).
Superimposed on this fluctuation is a linear
increase of the temperature by typically
$\approx0.40 \pm 0.01^o$C per century
ever since reliable data is available, \textit{i.e.}
since 1702 (errors are statistical only, one standard
deviation).
The world population has doubled from
1952 to 1990,
yet we see no statistically significant acceleration
of global warming in this period.
We conclude that the effect of humankind on global
warming up to 1990 is $0.0 \pm 0.1^o$C.
Therefore, contrary to popular belief,
the data support the view that human activity has
had no significant effect on global warming up
to the year 1990 covered by this study.
\end{abstract}


\section{Introduction}
We present a study of global warming based on the
data obtained at $13472$ weather stations covering
the period from 1702 to 1990. In particular
we are interested in finding the effect of human
activity on global warming. Since the world
population doubled from 1952 to 1990, we
search for any changes in the trend of global
warming between the first and second halves of
the 20'th century.

The data used in this study
is described in Section 2, the analysis is
presented in Section 3 and the conclusions are
collected in Section 4.

\section{The data}
The temperature data set used in this analysis is
\textquotedblleft{The Global Historical
Climatology Network (GHCN) version 2}"
released in 1997.\cite{GHCN}
This data
is a \textquotedblleft{comprehensive global surface
baseline climate data set designed for monitoring and
detecting climate change.}"
In particular we used the
\textquotedblleft{Adjusted Monthly Mean Temperature Data}"
(in file \textsf{v2.mean.adj})
and the \textquotedblleft{Raw Monthly Mean Temperature Data}"
(in file \textsf{v2.mean}). Each station is
labeled by a $12$ digit code.

\begin{figure}
\begin{center}
\vspace*{-6.0cm}
\scalebox{0.6}
{\includegraphics[0in,1in][8in,9.5in]{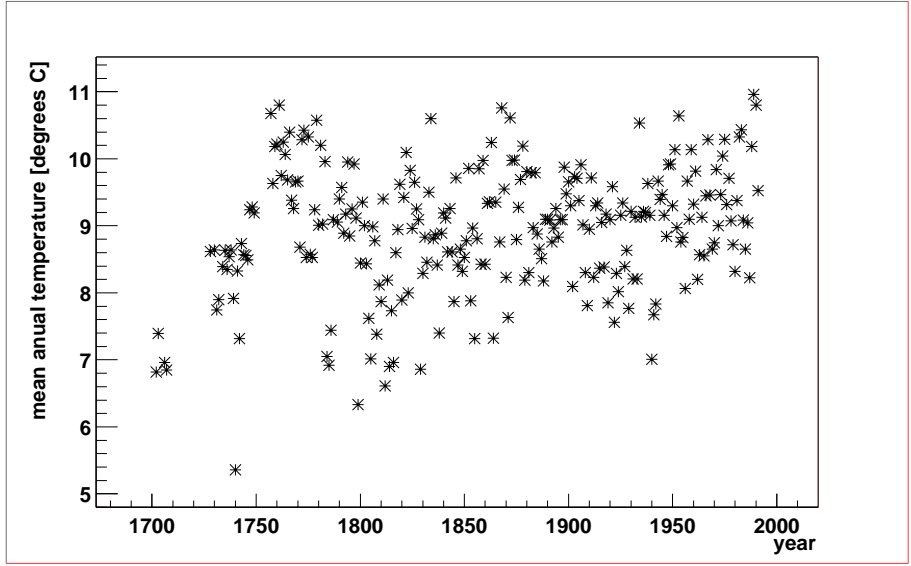}}
\vspace*{1.5cm}
\caption{Mean annual temperature of station
617103840000 in Germany. Raw data.
$a_1 = 0.27 \pm 0.08^o$C per century.}
\label{617103840000}
\end{center}
\end{figure}

\section{Discussion}
\begin{figure}
\begin{center}
\vspace*{-6.0cm}
\scalebox{0.6}
{\includegraphics[0in,1in][8in,9.5in]{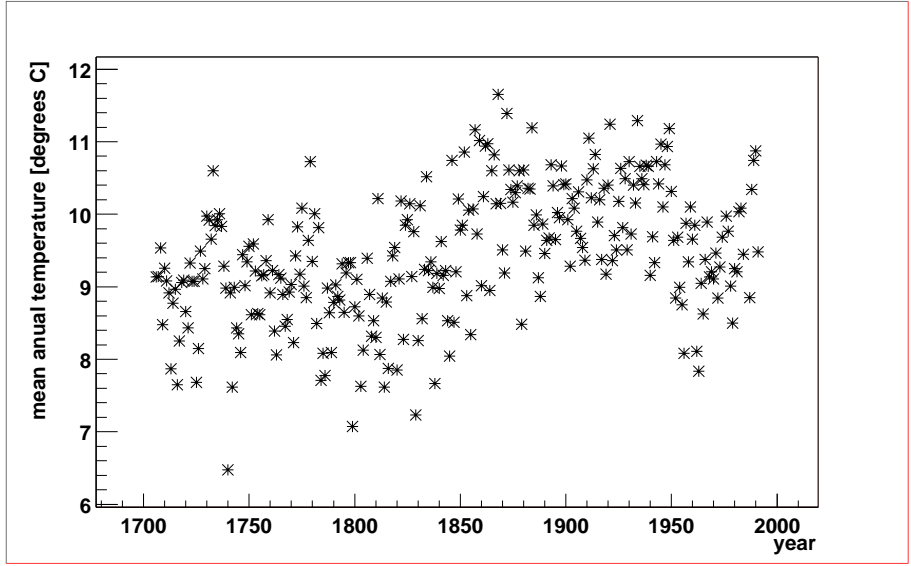}}
\vspace*{1.5cm}
\caption{Mean annual temperature of station
633062600000 in the Netherlands. Raw data.
$a_1 = 0.46 \pm 0.07^o$C per century.}
\label{633062600000}
\end{center}
\end{figure}

\begin{figure}
\begin{center}
\vspace*{-6.0cm}
\scalebox{0.6}
{\includegraphics[0in,1in][8in,9.5in]{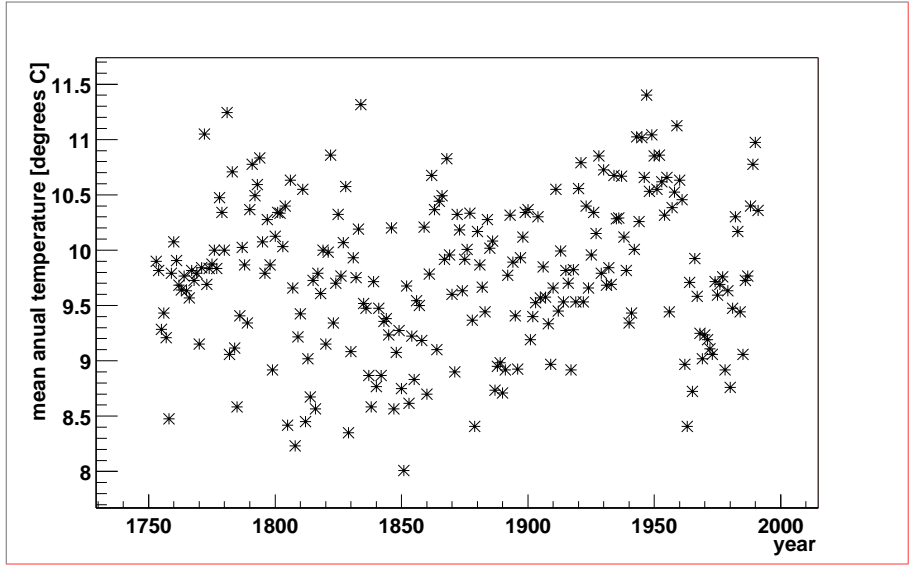}}
\vspace*{1.5cm}
\caption{Mean annual temperature of station
646067000000 in Switzerland. Raw data.
$a_1 = 0.12 \pm 0.06^o$C per century.}
\label{646067000000}
\end{center}
\end{figure}

\begin{figure}
\begin{center}
\vspace*{-7.0cm}
\scalebox{0.6}
{\includegraphics[0in,1in][8in,9.5in]{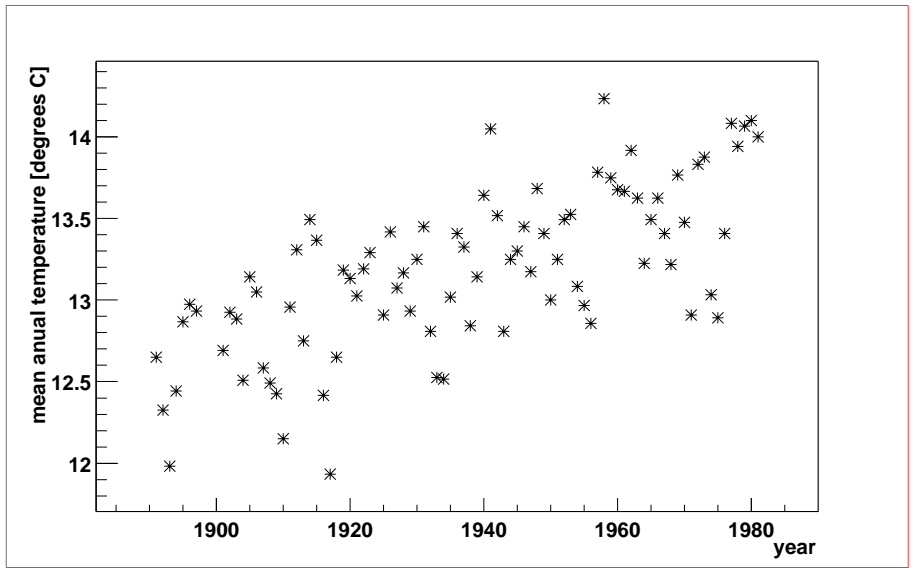}}
\vspace*{1.5cm}
\caption{Mean annual temperature of station
306840710000 in Quito, Ecuador. Raw data.
$a_1 = 1.3 \pm 0.2^o$C per century.}
\label{306840710000}
\end{center}
\end{figure}

We define a \textquotedblleft{valid year}" for a particular
station as a year that has all twelve months with valid mean
temperatures. We average these twelve monthly temperatures
to obtain the \textquotedblleft{mean annual temperature}"
for that station.
In Figures \ref{617103840000}, \ref{633062600000}
and \ref{646067000000} we present
the mean annual temperatures $T_i$ of the three oldest stations
using the \textquotedblleft{Raw Monthly Mean Temperature Data}"
set.
In Figure \ref{306840710000} we present the corresponding
temperatures of our home city Quito. We fit these
temperatures to a straight line. The slopes
$a_1$ for the period covered by the data are shown
in the figure captions. All errors in this article
are statistical and one standard deviation or $68\%$ confidence.
The high slope of the temperature measured in Quito
may be due to the fact that the city grew around
the weather station.

\begin{figure}
\begin{center}
\vspace*{-7.0cm}
\scalebox{0.6}
{\includegraphics[0in,1in][8in,9.5in]{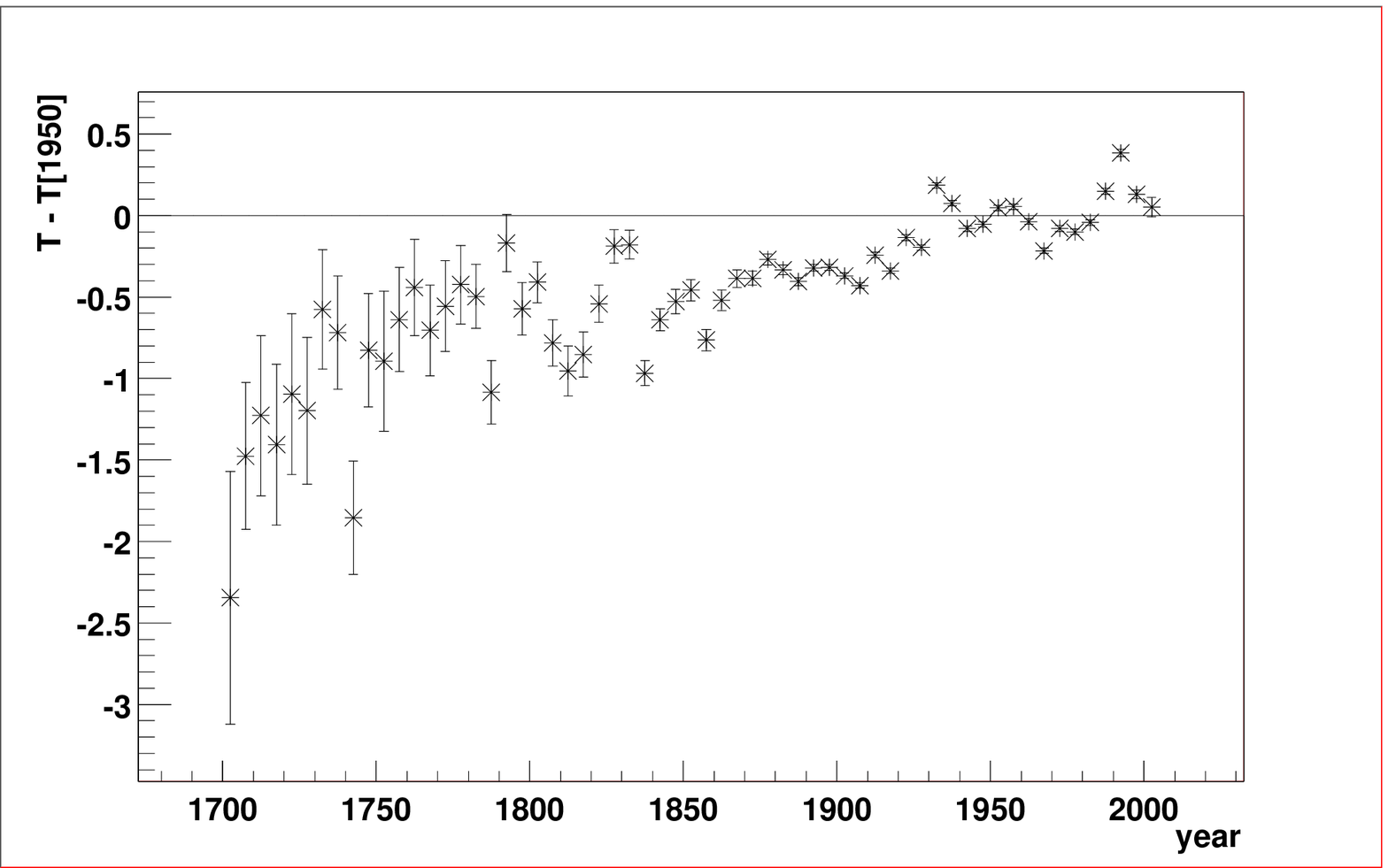}}
\vspace*{1.5cm}
\caption{Temperature relative to 1950 averaged over
5 year bins and averaged over the 456
stations satisfying the selection criteria described in the text.
Raw data. Errors are statistical,
one standard deviation. $a_1=0.400 \pm 0.009^o$C per century.
See text for details.}
\label{global_warming_raw}
\end{center}
\end{figure}

\begin{figure}
\begin{center}
\vspace*{-7.0cm}
\scalebox{0.6}
{\includegraphics[0in,1in][8in,9.5in]{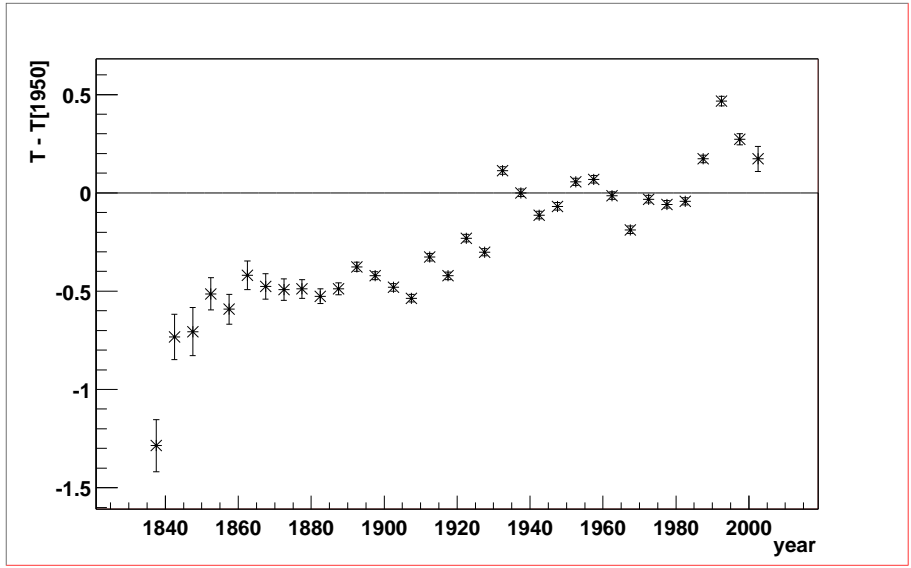}}
\vspace*{1.5cm}
\caption{Temperature relative to 1950 averaged over
5 year bins and averaged over the 377
stations satisfying the criteria described in the text.
Adjusted data. Errors are statistical,
one standard deviation. $a_1=0.582 \pm 0.010^o$C per century.
See text for details.}
\label{global_warming_adjusted}
\end{center}
\end{figure}

Figure \ref{global_warming_raw} presents global warming at-a-glance.
This figure was obtained as follows.
We consider stations that have the first valid year less
than or equal to 1900 and the last valid year greater than
or equal to 1990.
Furthermore we require at least 85 valid years between
1900 and 1990 inclusive.
We chose these stations to be able
to make meaningfull comparisons of global warming
between the first and second halves of the 20'th century.
We define the \textquotedblleft{reference temperature}"
of a station to be the average temperature of
valid years between 1945 and 1955 inclusive (so at least
one valid year in this interval is required).
In Figure \ref{global_warming_raw} we present the
difference between the average temperature per 5 year bin
and the reference temperature, averaged over all stations in the
\textquotedblleft{Raw Monthly Mean Temperature Data}" set
passing the selection criteria.
In Figure \ref{global_warming_adjusted} we present the
corresponding information for the
\textquotedblleft{Adjusted Monthly Mean Temperature Data}" set.

From these figures we see that the global temperature has
been increasing since the earliest measurements in 1702.
We see no acceleration of global warming in the second
half of the 20'th century compared to the first half.

The data shown in Figures \ref{global_warming_raw}
and \ref{global_warming_adjusted} are dominated by
stations in the USA as can be seen in Figure \ref{USA}.
However, the general conclusions, \textit{i.e.}
warming since the earliest measurements and no
acceleration of the warming in the second half of the
20'th century, can be observed elsewhere as shown
in Figures \ref{Germany} to \ref{Sweeden}.

\section{Conclusions}
\begin{figure}
\begin{center}
\vspace*{-7.0cm}
\scalebox{0.6}
{\includegraphics[0in,1in][8in,9.5in]{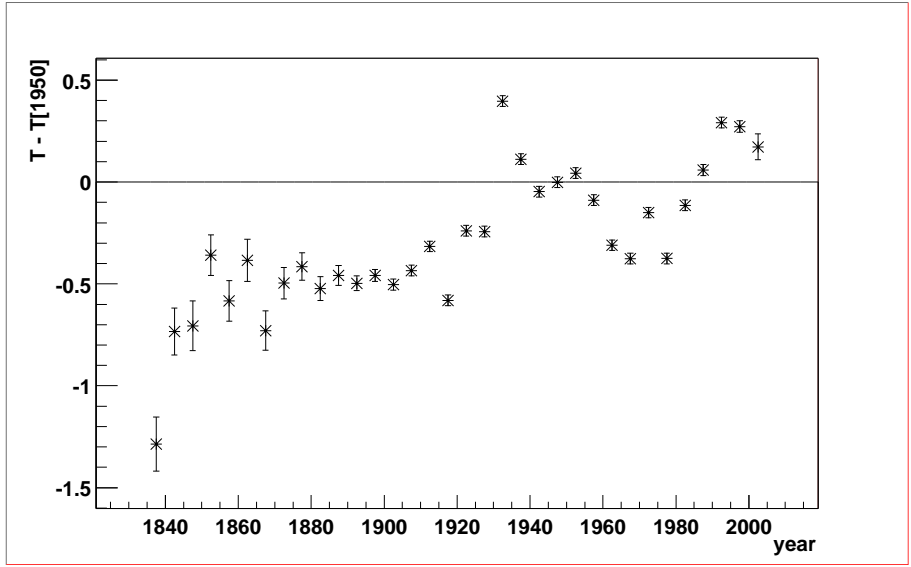}}
\vspace*{1.5cm}
\caption{Temperature relative to 1950 averaged over
5 year bins and averaged over the 247
stations in the USA satisfying the criteria described in the text.
Adjusted data. $a_1=0.462 \pm 0.013^o$C per century.}
\label{USA}
\end{center}
\end{figure}

\begin{figure}
\begin{center}
\vspace*{-7.0cm}
\scalebox{0.6}
{\includegraphics[0in,1in][8in,9.5in]{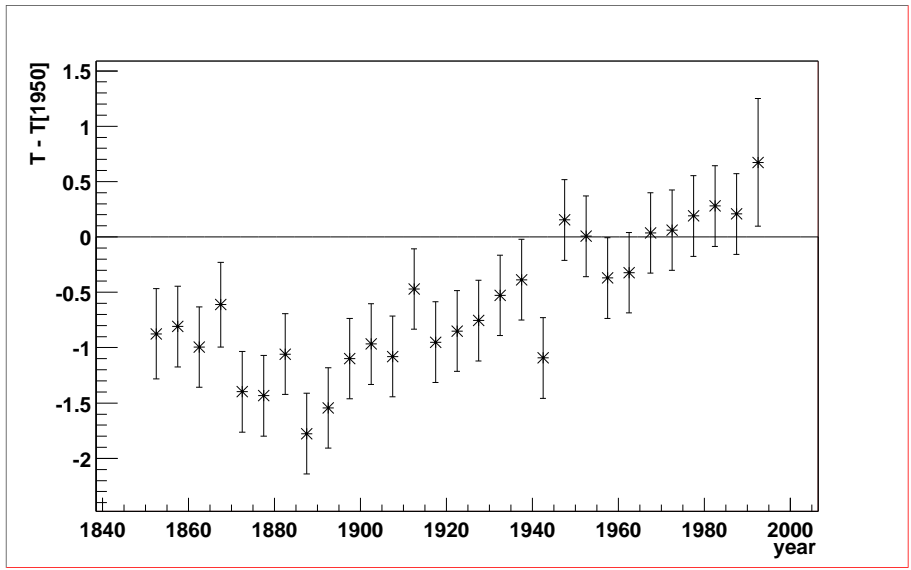}}
\vspace*{1.5cm}
\caption{Temperature relative to 1950 averaged over
5 year bins and averaged over the 2
stations in Germany satisfying the criteria described in the text.
Adjusted data. $a_1=1.18 \pm 0.13^o$C per century.}
\label{Germany}
\end{center}
\end{figure}

\begin{figure}
\begin{center}
\vspace*{-7.0cm}
\scalebox{0.6}
{\includegraphics[0in,1in][8in,9.5in]{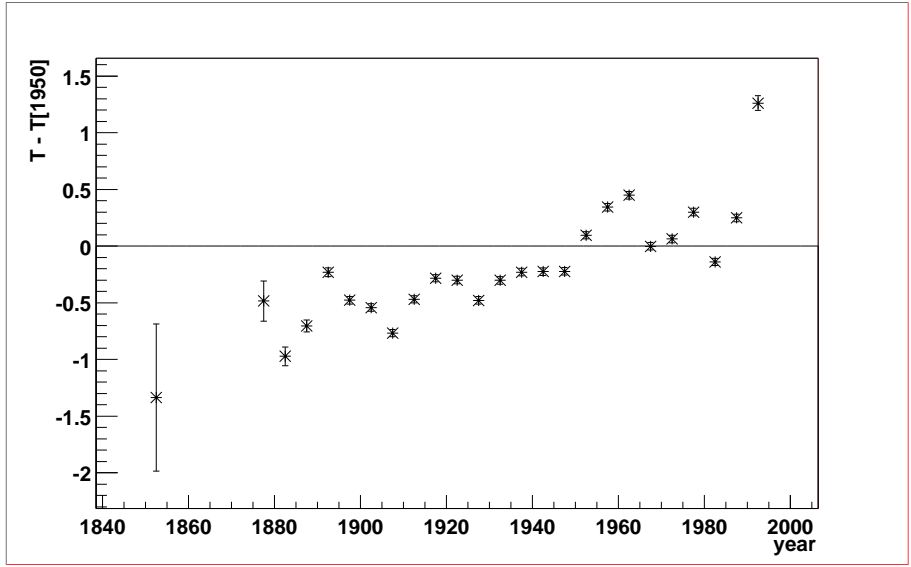}}
\vspace*{1.5cm}
\caption{Temperature relative to 1950 averaged over
5 year bins and averaged over the 75
stations in Japan satisfying the criteria described in the text.
Adjusted data. $a_1=0.987 \pm 0.022^o$C per century.}
\label{Japan}
\end{center}
\end{figure}

\begin{figure}
\begin{center}
\vspace*{-7.0cm}
\scalebox{0.6}
{\includegraphics[0in,1in][8in,9.5in]{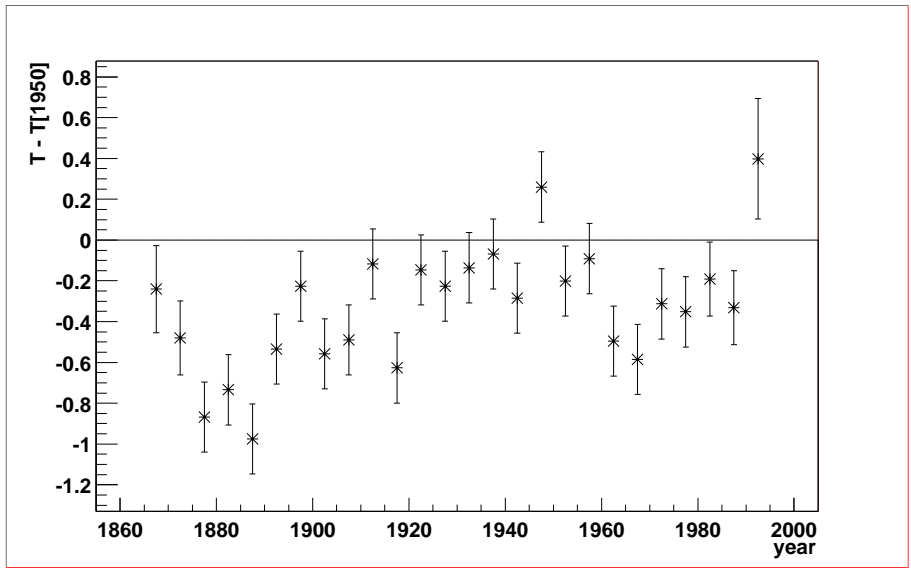}}
\vspace*{1.5cm}
\caption{Temperature relative to 1950 averaged over
5 year bins and averaged over the 3
stations in the United Kingdom
satisfying the criteria described in the text.
Adjusted data. $a_1=0.41 \pm 0.08^o$C per century.}
\label{UK}
\end{center}
\end{figure}

\begin{figure}
\begin{center}
\vspace*{-7.0cm}
\scalebox{0.6}
{\includegraphics[0in,1in][8in,9.5in]{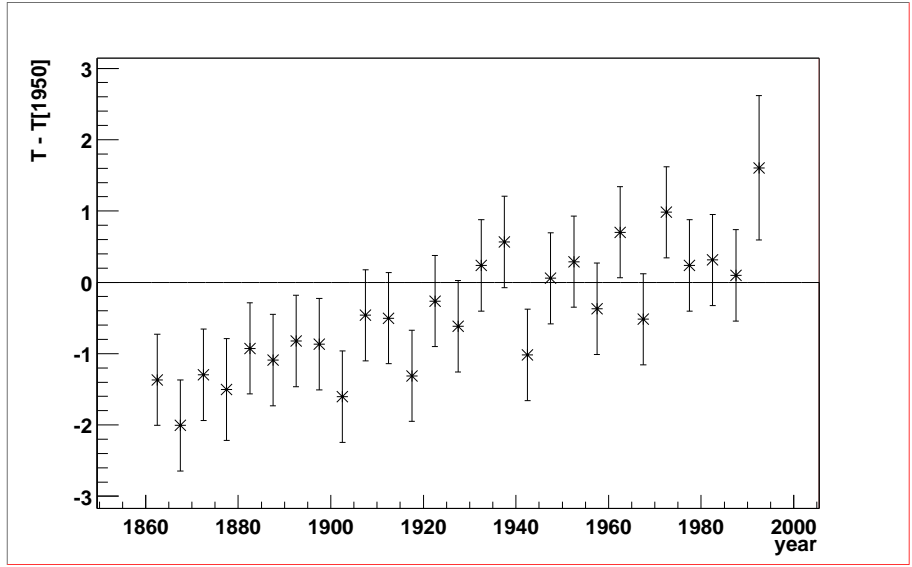}}
\vspace*{1.5cm}
\caption{Temperature relative to 1950 averaged over
5 year bins of the single
station in Sweeden satisfying the criteria described in the text.
Adjusted data. $a_1=1.75 \pm 0.30^o$C per century.}
\label{Sweeden}
\end{center}
\end{figure}

\begin{figure}
\begin{center}
\vspace*{-7.0cm}
\scalebox{0.6}
{\includegraphics[0in,1in][8in,9.5in]{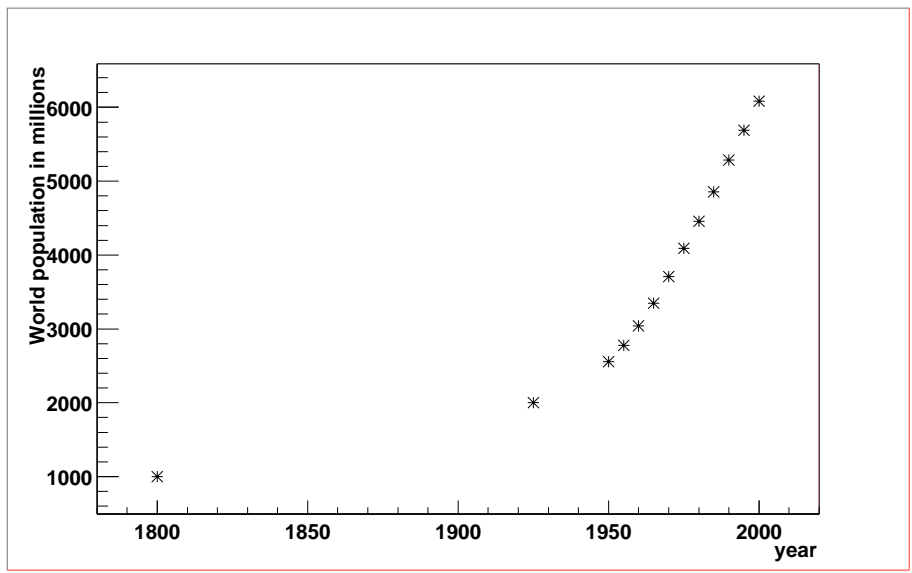}}
\vspace*{1.5cm}
\caption{World population in millions (from data
collected by the United Nations, except for the first
two points which are estimates).}
\label{pobl}
\end{center}
\end{figure}

The mean annual temperature of a station
fluctuates from year to year by typically
$\approx \pm 0.6^o$C (one standard deviation).
Superimposed on this fluctuation is a linear
increase of temperature by typically
$\approx 0.40 \pm 0.01^o$C per century
ever since reliable data is available, \textit{i.e.}
since 1702,
as shown in Figure \ref{global_warming_raw}.
All errors are statistical only, one standard
deviation.
The world population has \textbf{doubled} from
1952 to 1990 as seen in Figure \ref{pobl},
yet we see no statistically significant acceleration
of global warming in that period: see Figures
\ref{global_warming_raw} to \ref{Sweeden}.
We conclude that the effect of humankind on global
warming up to 1990 is $0.0 \pm 0.1^o$C.
Therefore, contrary to popular belief,
the data support the view that human activity has
had no significant effect on global warming up
to the year 1990 covered by this study.

\end{document}